\documentclass[12pt,preprint]{aastex}
\slugcomment{Astrophys. J., in press}

\shorttitle{Savin \& Laming}
\shortauthors{Uncertainties in Dielectronic Recombination}

\begin{document}

\title{Uncertainties in Dielectronic Recombination Rate Coefficients:
Effects on Solar and Stellar Upper Atmosphere Abundance Determinations}

\author{Daniel Wolf Savin}
\affil{Columbia Astrophysics Laboratory, Columbia University, \\ New York,
NY 10027}
\email{savin@astro.columbia.edu}
\author{J. Martin Laming}
\affil{E. O. Hulburt Center for Space Research, US Naval Research
Laboratory,
Code 7674L, Washington, DC 20375}
\email{jlaming@ssd5.nrl.navy.mil}

\begin{abstract}

We have investigated how the relative elemental abundances inferred
from the solar upper atmosphere are affected by uncertainties in the
dielectronic recombination (DR) rate coefficients used to analyze the
spectra.  We find that the inferred relative abundances can be up to a
factor of $\approx 5$ smaller or $\approx 1.6$ times larger than those
inferred using the currently recommended DR rate coefficients.  We have
also found a plausible set of variations to the DR rate coefficients
which improve the inferred (and expected) isothermal nature of solar
coronal observations at heights of $\gtrsim 50$ arcsec off the solar
limb.  Our results can be used to help prioritize the enormous amount
of DR data needed for modeling solar and stellar upper atmospheres.
Based on the work here, our list of needed rate coefficients for DR
onto specific isoelectronic sequences reads, in decreasing order of
importance, as follows: O-like, C-like, Be-like, N-like, B-like,
F-like, Li-like, He-like, and Ne-like.  It is our hope that this work
will help to motivate and prioritize future experimental and
theoretical studies of DR.

\end{abstract}

\keywords{atomic data -- atomic processes -- stars: abundances -- stars:
coronae -- Sun: abundances -- Sun: corona -- Sun: transition region}

\section{Introduction}
\label{sec:Intro}

It is now generally accepted that the elemental composition of the
solar wind is different from that of the photosphere.  Three decades of
research have shown that in the slow speed solar wind (which is
believed to originate from quiet coronal regions) the abundance ratio
of low first ionization potential (FIP) elements (FIP $<10$ eV)
relative to higher FIP elements is larger than it is in the
photosphere.  The observed FIP enhancement is roughly a factor of
four.  With the advent of the Ulysses spacecraft in its polar orbit, it
has become clear that in the fast solar wind (which emanates from
coronal holes) the enhancement of the low FIP elements is certainly
less than two, and is perhaps consistent with the solar photospheric
abundance pattern \citep{vonsteiger97,feldman00}.

Similar phenomena are observed in spectroscopic studies of the solar
corona and transition region, which we will refer to as the solar upper
atmosphere. Mapping the FIP enhancement factor from the solar
wind, through the corona and transition region, and into the
photosphere (where by definition the factor is 1) is an important area
of research for understanding the formation of the solar wind.  FIP
factors for the solar wind are determined from in situ particle
measurements and believed to be robust.  FIP factors for the solar
corona and transition region are inferred from spectroscopic
observations and are sensitive to a number of uncertainties.  Here we
will explore how uncertainties in the dielectronic recombination
(DR) rate coefficients limit our ability to
infer FIP factors in the solar upper atmosphere.

Of the many ionization and recombination rate coefficients which go
into ionization balance calculations for solar and stellar coronae
(i.e., electron-ionized plasmas), the high temperature DR rate
coefficients are believed to be the most uncertain
\citep{Arna92a,mazzotta98}.  Unlike most other atomic processes (e.g.,
ionization and excitation) where the direct contribution dominates the
process, for DR is solely a resonant process.  These resonances are
doubly-excited, intermediate states which are highly correlated which
makes the calculations theoretically and computationally challenging.

Especially important are the high temperature DR rate coefficients for
ions with partially filled $L$-shells. The astrophysical implications
of these uncertainties are poorly understood.  Investigating these
implications is important because DR is the dominant electron-ion
recombination process for most ions in cosmic plasmas.

Typically, inferences of the FIP effect from the solar corona are made
from spectra that exhibit lines from several charge states of the same
element \citep[see e.g.,][]{malinovsky73,laming95,white00}. In such
cases any uncertainties in the DR rate coefficients and the ionization
balance calculations essentially cancel out. This is because an error
that might increase the fraction of ions in a particular charge state,
does so at the expense of ionic fractions in neighboring charge
states.  If one observes a whole series of charge states to evaluate
the FIP effect \citep[e.g., Fe {\small IX, X, XI, XII, XIII, XIV, XV,
and XVI} in the case of][]{white00}, these problems are to a large
extent obviated.

Similar procedures are difficult to follow at solar transition region
temperatures where fewer elements have substantial series of observable
charge states.  The situation is compounded by the temperature
dependence of the emission measure, as discussed in
\S~\ref{subsec:SolarDisk}, and by the trend in solar physics
instrumentation towards spectrometers with ever more limited bandpass
while spatial and spectral resolution and sensitivity improve.  For
example, the spectrometers on the Solar Orbiting Heliospheric
Observatory ({\it SOHO}) such as the Coronal Diagnostic Spectrometer
\citep[CDS;][]{harrison95,harrison97} and the Solar Ultraviolet
Measurements of Emitted Radiation
\citep[SUMER;][]{wilhelm95,wilhelm97,lemaire97} must scan their
detectors across the bandpass to build up a full spectrum.  This limits
their utility in observing time varying plasmas. The Extreme
Ultraviolet Imaging Spectrometer \citep{culhane01} will use multilayer
coated gratings, which will further restrict the bandpass. The proposed
wavelength ranges, $\approx 170-210$ and $\approx 250-290$ \AA\ will
provide good coverage in strong lines of Fe {\small IX - XVI} and
\ion{Fe}{24}, but are less satisfactory for ions of other
elements.  For these reasons accurate DR rate coefficients for the
few lines that will be studied in current and future solar satellite
missions have become more essential than ever.

Here we investigate the extent to which FIP enhancement factors,
inferred using individual line ratios, can be affected by uncertainties
in the DR rate coefficients used in the data analysis.  We hope
to motivate further experimental and theoretical studies of DR.
We begin by reviewing in \S~2 the status of the theoretical DR results
currently used in ionization balance calculations.  We focus primarily
on the rate coefficients for DR onto the $L$-shell ions \ion{Ne}{6}
through \ion{Ne}{8}, \ion{Mg}{6} through \ion{Mg}{9}, \ion{Si}{6}
through \ion{Si}{12}, and \ion{S}{9} and \ion{S}{10}.  Lines from these
ions are commonly observed from solar and stellar coronae.  In \S~3 we
discuss how relative abundances are determined for solar and stellar
upper atmospheres and the current observational status of relative
abundance determinations.  We describe in \S~4 the ionization balance
calculations, how the calculations are sensitive to our estimated
uncertainties in the DR rate coefficients, and present some
implications.  We conclude in \S~5 by making some specific
recommendation for future directions in theoretical and experimental
studies of DR.

\section{Dielectronic Recombination (DR)}
\label{sec:DR}

DR is a two-step process.  It begins when a free electron
collisionally excites an ion and is simultaneously captured (i.e.,
dielectronic capture).  Core excitations of the ion can be labeled $Nlj
\rightarrow N^\prime l^\prime j^\prime$.  Here $N$ is the principal
quantum number of the core electron, $l$ is its angular momentum, and
$j$ the total angular momentum of the core electron.  In general, the
most important DR channels are those via $\Delta N \equiv N^\prime -N =
0$ or 1 core excitations.  The incident electron is captured into some
$nl^{\prime \prime}$ Rydberg level forming any one of an infinite
number of intermediate doubly-excited states of the recombining ion.
These states can either autoionize or radiatively stabilize, the
latter of which completes the DR process.  Energy conservation
requires $E_k=\Delta E - E_b$, where $E_k$ is the kinetic energy of the
incident electron, $\Delta E$ the excitation energy of the core
electron, and $E_b$ the binding energy released when the free electron
is captured.  $\Delta E$ and $E_b$ are quantized, making DR is a
resonant process.  The DR rate coefficient represents the convolution
of these resonances with a Maxwellian electron energy distribution.

Here we are interested in low-electron-density, zero-field DR
rate coefficients.  Ionization balance calculations generally assume
that electron densities are low enough that collisional ionization of
the weakly bound electron captured in the DR process is unimportant.
These calculations also assume that any external
electric and magnetic fields are too weak to affect the DR rate
coefficient, and that metastable populations in all ions are
insignificant.

\subsection{Overview of Theory and Experiment}
\label{subsec:overview}

Reliable calculations of DR are extremely challenging theoretically and
computationally.  In contrast to most other atomic processes (e.g.,
ionization and excitation) where the direct contribution dominates the
process, DR is solely a resonant process.  These resonances are
doubly-excited, intermediate states which  are highly correlated.
These states greatly complicate calculations, since an accurate
treatment of electron correlation is required \citep{Zong97a}.
Calculations also require accounting for an infinite number of states.
This is clearly impossible using a finite basis expansion.
Approximations must be made to make the calculations tractable
\citep{Hahn93a}.

Many different theoretical techniques have been used to calculate DR
rate coefficients for plasma modeling.  In the past, semi-empirical
expressions such as the \citet{Burg65a} formula along with modified
versions by \citet{Burg76a} and \citet{Mert76a} were derived to
calculate DR rate coefficients.  More recently, a number of
sophisticated theoretical approaches have been developed, among them
configuration-averaging \citep{Grif85a}; single-configuration
$LS$-coupling \citep{McLa84a}; quantum defect \citep{Bell85a};
intermediate-coupling \citep{Badn89b}; nonrelativistic,
multiconfiguration Hartree-Fock \citep[MCHF;][]{Nils86a,Schi98a};
semi-relativistic, multiconfiguration Breit-Pauli
\citep[MCBP;][]{Badn86a}; fully relativistic, multiconfiguration
Dirac-Fock \citep[MCDF;][]{Chen85a}; and relativistic many-body
perturbation theory \citep[RMBPT;][]{Zong97a} methods.  Other methods
include the Hebrew University Lawrence Livermore Atomic Code (HULLAC)
which uses a relativistic, multiconfiguration, parametric potential
method \citep{Mitn98a} and $R$-matrix methods where radiative
recombination (RR) and DR are treated in a unified manner in the
close-coupling approximation \citep{Naha94a,Robi95a}.  However, the
result of all these different theoretical techniques and required
approximations are DR rate coefficients which often differ by factors
of $\sim 2$ to 4 or more
\citep{Arna92a,Savi97a,Savi99a,Savi99b,Savi00a}.

Cosmic plasmas are most commonly modeled using the recommended DR rate
coefficients of \citet{Aldr73a}, \citet{Shul82a},
\citet{Nuss83a,Nuss84a,Nuss86a,Nuss87a}, \citet{arnaud85},
\citet{Land91a}, \citet{Arna92a}, and \citet{mazzotta98}.  Essentially
none of these DR rate coefficients have been calculated using
state-of-the-art techniques (i.e. MCHF, MCBP, MCDF, RMBPT, HULLAC, or
$R$-matrix techniques which include the spin-orbit interaction).  The
vast majority of these have been calculated using single-configuration
pure $LS$-coupling, semi-empirical formulae, or isoelectronic
interpolations.

It is important to carry out calculations using techniques more
sophisticated than $LS$-coupling because $LS$-coupling calculations are
known not to include all possible autoionization levels contributing to
the DR process, due to not including the spin-orbit
interaction.  As a result, such calculations provide only a lower limit
for the DR rate coefficient \citep{Badn88a,Gorc96a}.  For example, for
lithiumlike ions, Griffin et al.\ (1985) and \citet{Beli87a} discussed
how $LS$-coupling accounts for only two-thirds of all possible $\Delta
N=0$ recombining channels.  Intermediate coupling calculations yield a
DR rate coefficient for lithiumlike C~IV which is 50\% larger than the
$LS$-coupling result.  Recent ion storage ring measurements and
relativistic many-body perturbation calculations have verified the
importance of these $LS$-forbidden autoionizing resonances
\citep{Mann98a}.  As for the various semi-empirical formulae,
\citet{Savi99b} showed that a priori it is not possible to know which
of the semi-empirical formulae will yield a result close to the true DR
rate coefficient and which will be off by a factor of 2.

Due to the complexity of the theoretical descriptions for the DR
process, it is almost impossible to say {\it a priori} which
approximations in the calculations are justified and which are not.
Laboratory measurements are needed to ensure that even state-of-the-art
techniques produce reliable results.  For example, initial \ion{Fe}{16}
MCBP results were a factor of $\approx 2$ larger than ion storage ring
measurements \citep{Link95a}.  This discrepancy was later resolved
\citep{Gorc96a}; but without laboratory measurements, the error in the
theory would probably have gone undiscovered for many years.

A related difficulty is that, for many of the ions and DR resonances
important in electron-ionized cosmic plasmas, there are very few
experimental techniques capable of carrying out measurements at the
accuracy needed to provide benchmarks useful for the theorists.
Tokamak and theta pinch measurements suffer from factor of 2
uncertainties \citep{Grie88a}.  Crossed electron-ion beams techniques
\citep{Mull87a,Savi96a} have been limited in the energy range
accessible as well as by low signal rates.  Single-pass, merged
electron-ion beam techniques \citep{Ditt87a,Ande90a,Ande92a,Sche94a}
can cover a wider energy range but also have low signal rates.  Due to
a combination of poor statistics and poorly controlled external
electric and magnetic fields, it is difficult to use most of these
crossed- and merged-beams measurements to benchmark zero-field DR
calculations.  Using either technique it is also difficult to determine
the ion beam metastable population.  Hence these techniques cannot be
used to carry out absolute measurements for many ions with partially
filled shells \citep{Badn91a}.

The two state-of-the-art experimental techniques for studying DR are
electron beam ion traps (EBITs) and heavy-ion storage rings with a
merged electron-ion beam interaction region.  Both techniques store the
ions long enough for all metastable states to decay radiatively to the
ground state.  Storage ring techniques measure absolute DR resonance
strengths.  EBITs produce relative resonance strength measurements
which can then be normalized to RR or electron impact excitation
theory.

Until recently EBIT measurements were essentially limited to studying
only closed shell systems \citep{Beie92a,Smit00a}.  This has changed
with the recent studies of DR onto \ion{Fe}{21} $-$ 
\ion{Fe}{24} \citep{Gu99a,Gu00a}.  EBITs, however,
are still not capable of reliably studying DR for collision
energies $\lesssim 670$~eV \citep{Warg01a}.  These are the dominant DR
channels for most ions of second and third row elements, especially for
ions with partially filled shells.  Work is underway to overcome this
limitation.

The majority of the storage ring measurements for astrophysically
important ions have been carried out using TSR and CRYRING
\citep{Mull95a}.  Measurements of DR resonance strengths and energies
have been reported using TSR from 0 to $\approx 2200$~eV
\citep{Kenn95a,Link95a} and using CRYRING from 0 to $\approx 50$~eV
\citep{DeWi95a,DeWi96a}.  Storage rings are the optimal technique for
studying DR resonances at energies not accessible to EBITs.  This makes
storage rings particularly well suited for studying DR for ions of
second and third row elements.  The external fields in the merged beams
section are also extremely well controlled which allows reliable,
essentially zero-field DR measurements to be carried out
\citep{Savi00b}.  One limitation, though, is that ions with a
charge-to-mass ratios $q/m \lesssim 0.1$ cannot be accelerated in
CRYRING and TSR to a velocity where electron capture from the rest gas
in the ring becomes negligible.  The resulting poor signal-to-noise
ratio prevents measurements of DR for such ions
\citep{Schi98a,Wolf99a}.

To date, EBIT and storage ring DR measurements have been limited in the
isoelectronic sequences studied and the energy ranges covered.  Few
data sets exist which are comprehensive enough to benchmark  high
temperature DR theory.  And for those ions studied over a wide enough
energy range, the benchmarked theory has not then been used to
calculate the DR rate coefficients for all astrophysically important
ions isoelectronic to the measured ion.  As a result, only a very small
fraction of the DR rate coefficients used for modeling electron-ionized
cosmic plasmas have come from benchmarked state-of-the-art theory.

\subsection{Uncertainties in the Recommended DR Rate Coefficients}
\label{subsec:Uncertainties}

Here we are interested in how uncertainties in the published DR rate
coefficients affect the ionization balance calculations used to analyze
solar and stellar spectra.  The ions of interest are listed in
\S~\ref{sec:Intro}.  However, the calculated abundance for a given ion
is affected by the atomic rate coefficients for the ions one lower and
higher in charge state (and less so by ions two or more lower and
higher).  To account for this, we extend the list of ions of interest
by one charge state on both the low and high end for each element
considered.  Below we review and estimate uncertainties in the
published high temperature rate coefficients for DR onto
\ion{Ne}{5} $-$ \ion{Ne}{9}, \ion{Mg}{5} $-$ \ion{Mg}{10},
\ion{Si}{5} $-$ \ion{Si}{13}, and \ion{S}{8} $-$ \ion{S}{11}.

In the absence of measurements, the best way to estimate the
uncertainty in the theoretical DR rate coefficients is to compare
calculations along an isoelectronic sequence.  This yields a
conservative estimate.  \citet{Savi99a} showed that published rate
coefficients do not necessarily give reliable upper and lower limits
for the range in which the true DR rate coefficient lies.  The true DR
rate coefficient may lie outside of these limits.  Hence our approach
here may, in fact, underestimate the uncertainties in the DR rate
coefficients

We compare the range of calculated DR rate coefficients to the
recommended DR rate coefficients from \citet{mazzotta98} in order to
estimate the uncertainties in the ``state-of-the-art'' for ionization
balance calculations.  Their DR rate coefficients are meant for use at
temperatures between $10^4$ and $10^9$~K.  Wherever possible we
also use laboratory measurements to benchmark theory.  Because
experimental data often do not exist for the ions we are interested, we
take into account measurements carried out on those isoelectronic ions
which are closest in atomic number to the ions of interest.  We also
only take into account measurements of those resonances which are
relevant to the DR rate coefficients of interest here.  In our
discussion below, the predicted temperature of formation for an ion in
an electron-ionized plasma is taken from the results of Mazzotta et
al.  Also, estimated uncertainties for DR rate coefficients onto a
given ion are quoted for temperatures near where the ion peaks in
fractional abundance in an electron-ionized plasma.

Based on the comparisons described below, for the ions of interest we
have put together a list of factors by which to scale the recommended
DR rate coefficients of \citet{mazzotta98}.  These scale factors are
listed in Table~\ref{tab:scalefactors}.  The $2^6 \times 3^3$ possible
combinations yield a total of 1728 sets of DR rate coefficients which
we will use in \S~\ref{sec:Effects}.

\subsubsection{Onto Heliumlike \ion{Ne}{9} and \ion{Si}{13}}
\label{subsubsec:He}

$LS$-coupling calculations were carried out by \citet{Jaco77b}, for
only \ion{Si}{13}, and by \citet{Roma88a}.  MCHF calculations were
published by \citet{Nils86a} and \citet{Kari89a}.  \citet{Chen86a} has
published MCDF rate coefficients.  The results of Chen, Nilsen,
Romanik, and Karim \& Bhalla are in excellent agreement while the
results of Jacob et al.\ are a factor of $\approx 2$ smaller.
\citet{mazzotta98} use the rate coefficients of Chen.

EBIT and electron beam ion source studies of DR
have been carried out for a number of heliumlike
ions.  The most relevant measurements for our comparison here have been
on \ion{Ne}{9} (Wargelin, Kahn, \& Beiersdorfer 2000) and \ion{Ar}{17}
\citep{Ali91a,Smit96a}.  There is an estimated total experimental
uncertainty of $\approx 20\%$ for these results.  Overall there is good
agreement between theory and experiment.  Calculations using the
technique of \citet{Kari89a} are in good agreement with the
\ion{Ar}{17} measurements of \citet{Ali91a}.  Calculations using the
technique of \citet{Chen86a} are in good agreement with the results on
\ion{Ar}{17} \citep{Smit96a}.  For \ion{Ne}{9}, for the sum of the
measured \ion{Ne}{9} $KLL$ resonances, experiment lies a factor of 1.16
below calculations using the technique of \citet{Kari89a}, a factor of
1.23 below calculations using the technique of \citet{Chen86a}, and a
factor of 1.32 below calculations using the technique of
\citet{Nils86a}.

Overall the calculations of \citet{Chen86a} agree with laboratory
results to within $\approx 20\%$.  We take this estimate as the
uncertainty in the recommended rate coefficients of \citet{mazzotta98}
for DR onto \ion{Ne}{9} and \ion{Si}{13}.

\subsubsection{Onto Lithiumlike \ion{Ne}{8}, \ion{Mg}{10}, and \ion{Si}{12}}
\label{subsubsec:Li}

$LS$-coupling DR calculations have been published by several different
workers.  \citet{Jaco77b,Jaco79a} and \citet{Roma88a} published
coefficients for \ion{Ne}{8}, \ion{Mg}{10}, and \ion{Si}{12}.
\citet{Rosz87a} gave rate coefficients for \ion{Ne}{8}.  MCDF
results were published by \citet{Chen91a} for \ion{Ne}{8} and
\ion{Si}{12}.  For reasons discussed in \S~\ref{subsec:overview},
we do not use the published $LS$ results to estimate the uncertainty in
the DR rate coefficients onto lithiumlike ions.

Measurements have been carried out for a number of
lithiumlike ions.  Storage ring measurements have been carried out on
\ion{Ne}{8} \citep{Zong98a} and \ion{Si}{12}
\citep{Kenn95,Kenn95a,Bart97a}.  Uncertainties were
typically $\approx 20\%$.

At the temperature of peak formation, the dominant DR channel for
lithiumlike ions up to \ion{Si}{12} is via $\Delta N=0$ core
excitations.  For \ion{Si}{12} the $\Delta N=0$ and $\Delta N=1$ ($N=2
\to N^\prime=3$) contributions are comparable \citep{Chen86a}.  For
$\Delta N=0$ DR onto \ion{Ne}{8} MCBP theory was $\approx 20\%$ below
the experimental data.  MCPB theory was $\approx 10-20\%$ below
experiment for $\Delta N=0$ DR onto \ion{Si}{12}.  For \ion{Si}{12}
$\Delta N=1$ DR via $N=2\to N^\prime=3$ core excitations, MCBP theory
was larger than experiment by $\approx 10-20\%$ for the $1s^2 3l
3l^\prime$ resonances.  For the $1s^2 3l nl^\prime \ (n\ge4)$
resonances in \ion{Si}{12}, theory lies slightly below experiment.
These comparisons suggest that the accuracy of the MCBP results is
$\approx \pm 20\%$.

None of the measured resonance strengths have been compared with
results from the theoretical techniques used to calculate the published
rate coefficients for \ion{Ne}{8}, \ion{Mg}{10}, and \ion{Si}{12}.
This makes it difficult to use the measurements to infer an uncertainty
in the recommended DR rate coefficients.  We try to do this indirectly
by comparing MCBP results with published MCDF results.  The MCDF
results of \citet{Chen91a} agree to within $\approx 10\%$ with the MCBP
rate coefficients of \citet{Badn89b} for \ion{O}{6} and of
\citet{Badn99a} for \ion{Ar}{16}.  Considering the comparisons between
the MCBP and MCDF results and given the estimated uncertainty in the
MCBP results, we therefore estimate that there may be an $\approx \pm
20\%$ uncertainty in the MCDF rate coefficients of \citet{Chen91a}.

\citet{mazzotta98} use the rate coefficients of \citet{Chen91a} and
interpolate isoelectronically for those ions which Chen did not
calculate.  Chen did not present results for $T \lesssim 10^5$~K.  As a
result the fitted rate coefficients of Mazzotta et al.\ do not have the
correct low temperature behavior.  This can readily be seen by plotting
their recommended rate coefficient for \ion{C}{4} and comparing it with
published \ion{C}{4} rate coefficients \citep{Schi01a}. This error in
the rate coefficients of Mazzotta et al.\ will affect ionization
balance calculations for photoionized plasmas but is expected to have
little effect on modeling electron-ionized plasmas.  Given the
estimated uncertainty in the results of Chen, we estimate that for
\ion{Ne}{8}, \ion{Mg}{10}, and \ion{Si}{12} there is an $\approx \pm
20\%$ uncertainty in the relevant DR rate coefficients of Mazzotta et
al.

\subsubsection{Onto Berylliumlike \ion{Ne}{7}, \ion{Mg}{9}, and \ion{Si}{11}}
\label{subsubsec:Be}

$LS$-coupling rate coefficients were published for \ion{Ne}{7}
\ion{Mg}{9}, and \ion{Si}{11} by \citet{Jaco77b,Jaco79a} and
\citet{Roma88a}.  Near the temperatures of peak formation, their high
temperature results are in good agreement.  MCBP results were
presented by \citet{Badn87a} for \ion{Ne}{7}, \ion{Mg}{9}, and
\ion{Si}{11}.  At the temperature of peak formation, the rate
coefficients of Jacobs et al.\ and Romanik are a factor of $\approx
1.4-1.6$ times larger than those of Badnell.

\citet{mazzotta98} use the rate coefficients of \citet{Badn87a} and
appear to interpolate along the isoelectronic sequence for those ions
which Badnell did not calculate.  Badnell does not present any results
for $T < 10^5$~K.  A comparison of the rate coefficients from Mazzotta
et al.\ with those from \citet{Roma88a} shows that the former do not
have the correct low temperature behavior.  This will be important in
modeling the ionization structure of photoionized plasmas but is not an
issue here.

In the absence of laboratory benchmarks, we use the various published
DR calculations to provide upper and lower limits for the DR rate
coefficients.  As discussed above, this is a conservative estimate of
the uncertainty.  We estimate that for \ion{Ne}{7}, \ion{Mg}{9}, and
\ion{Si}{11}, the uncertainty in the relevant DR rate coefficients of
\citet{mazzotta98} is +60\% and -0\%.

\subsubsection{Onto Boronlike \ion{Ne}{6}, \ion{Mg}{8}, and \ion{Si}{10}}
\label{subsubsec:B}

\citet{Jaco77b,Jaco79a} reported $LS$-coupling DR rate coefficients for
\ion{Ne}{6}, \ion{Mg}{8}, and \ion{Si}{10}.  \citet{Naha95a} used
$R$-matrix techniques to calculate RR+DR rate coefficients in
$LS$-coupling for \ion{Ne}{6}, \ion{Mg}{8}, and \ion{Si}{10}.
\citet{mazzotta98} recommends DR rate coefficients based on the
calculations of Nahar after apparently subtracting the theoretical RR
rate coefficient from her results.

Given the paucity of theoretical calculations and appropriate
laboratory measurements, it is difficult to estimate the uncertainty in
the theoretical DR rate coefficients for the ions of interest.  We
attempt to do this indirectly by comparing the results for DR onto the
isoelectronic ions \ion{C}{2}, \ion{N}{3}, and \ion{O}{4} for which a
number of different calculations exist.  $LS$-coupling rate
coefficients were reported by \citet{Jaco78a} and \citet{Rama89a}.
Fits to the results of \citet{Jaco78a} were reported by
\citet{Shul82a}.  $R$-matrix results using $LS$-coupling were presented
by \citet{Naha95a}.  \citet{Badn89a} carried out MCBP calculations and
\citet{Safr98a} carried out MCHF calculations.  The various theoretical
techniques used have not yet converged to the same rate coefficients.
\citet{mazzotta98} recommends the rate coefficients of
\citet{Naha95a}.  At the temperatures of peak formation for \ion{C}{2},
\ion{N}{3}, and \ion{O}{4}, the other theoretical rate coefficients are
a factor of $\approx 1.0-1.7$ larger than the recommended rate
coefficients. We estimate that for \ion{Ne}{6}, \ion{Mg}{8}, and
\ion{Si}{10}, the uncertainty in the relevant DR rate coefficients of
Mazzotta et al.\ is +70\% and -0\%.

\subsubsection{Onto Carbonlike \ion{Ne}{5}, \ion{Mg}{7}, \ion{Si}{9},
and \ion{S}{11}}
\label{subsubsec:C}

There have been very few calculations of high temperature DR for the
ions of interest here.  $LS$-coupling results have been presented for
\ion{Ne}{5}, \ion{Mg}{7}, \ion{Si}{9}, and \ion{S}{11} by
\citet{Jaco77b,Jaco79a}.  \citet{mazzotta98} use the rate coefficients
of Jacobs et al.\ scaled up by a factor of $\approx 1.6$.
This factor was apparently derived by scaling the isoelectronic
\ion{Fe}{21} results
of \citet{Jaco77a}, calculated using the same technique as
\citet{Jaco77b,Jaco79a}, to the recommended DR rate coefficient of
\citet{Arna92a}.

Considering how few relevant calculations exist, it is difficult to
estimate the uncertainty in the DR rate coefficients for the ions of
interest here.  We attempt to do this indirectly using theoretical
results for high temperature DR onto \ion{N}{2} and \ion{O}{3}.
\citet{Jaco78a} presented $LS$-coupling results for \ion{N}{2} and
\ion{O}{3} which were fitted by \citet{Shul82a}.  $R$-matrix results in
$LS$-coupling were calculated by \citet{Naha97a} for \ion{N}{2} and by
\citet{Naha99a} for \ion{O}{3}.  MCBP and $LS$ rate coefficients were
calculated by \citet{Badn89b} and \citet{Rosz89a}, respectively, for
\ion{O}{3}.  Note that there is an apparent error in Table 1 of
\citet{Rosz89a}.  We had to reduce the DR rate coefficients by an order
of magnitude in order for the tabulated rate coefficients to match
those shown in Figures 1 and 2 of his paper.  Near the high temperature
DR peak for \ion{N}{2}, the rate coefficient of \citet{Naha97a} is a
factor of $\approx 2.2$ smaller than that of Jaocbs et al. For
\ion{O}{3} near this peak, the rate coefficient of Roszman is a factor
of $\approx 1.2$ times larger than that of Jacobs et al.  The rate
coefficients of Badnell \& Pindzola and Nahar \& Pradhan are,
respectively, $\approx 1.25$ and $\approx 1.7$ smaller than the results
of Jacobs et al.

For DR onto \ion{Ne}{5}, \ion{Mg}{7}, \ion{Si}{9}, and \ion{S}{11}, we
use the recommended rate coefficients of \citet{mazzotta98}.  A
reduction in these rate coefficients by 38\% brings their results into
agreement with the original results of Jacobs et al.  Considering the
comparison of the various theoretical DR rate coefficients of
\ion{N}{2} and \ion{O}{3}, a reduction in the rate coefficients of
Mazzotta et al.\ by 69\% represents our estimated lower limit to the
uncertainty in their recommended DR rate coefficients.

\subsubsection{Onto Nitrogenlike \ion{Mg}{6}, \ion{Si}{8}, and \ion{S}{10}}
\label{subsubsec:N}

We are aware only of the $LS$-coupling calculations by
\citet{Jaco77b,Jaco79a} for \ion{Mg}{6}, \ion{Si}{8}, and \ion{S}{10}.
\citet{mazzotta98} use the rate coefficients of Jacobs et al. For
\ion{S}{10}, Mazzotta et al.\ do not include the low temperature
results of Jacobs et al.  This is not an issue here but may be
important in photoionized plasmas.

The paucity of calculations for DR onto nitrogenlike ions makes it
difficult to estimate the uncertainty in the published theoretical rate
coefficients.  We attempt to do this indirectly using published rate
coefficients for DR onto the isoelectronic \ion{O}{2}.  High
temperature DR rate coefficients have been calculated using
$LS$-coupling by \citet{Jaco78a} and \citet{Tera91a}.  Fits to the
results of Jacobs et al.\ were published by \citet{Shul82a}.
\citet{Naha99a} published $R$-matrix rate coefficients (e.g. RR+DR)
using $LS$-coupling.  Intermediate coupling calculations have been
given by \citet{Badn89b} which were later improved upon by
\citet{Badn92a}.

The results of \citet{Jaco78a} and \citet{Badn89b} are in good
agreement.  However, \citet{Badn92a} recalculated the DR rate
coefficient taking into account correlation between the $n=2$ and $n=3$
shells.  His new rate coefficient is a factor of $\approx 1.6$ times
smaller than that of \citet{Jaco78a} and \citet{Badn89a}.
\citet{Tera91a} revised their published rate coefficients upward by
$\approx 20\%$ \citep{Badn92a}.  This revised result is a factor of
$\approx 2.2$ smaller than the results of \citet{Jaco78a}.  To
determine the DR rate coefficient of \citet{Naha99a}, we subtracted out
the estimated RR rate coefficient.  By extrapolating her recombination
rate coefficient at low temperatures, where DR is unimportant, we
estimated the RR rate coefficient at the relevant high temperatures.
The resulting DR rate coefficient lies a factor of $\approx 2.2$ below
that of Jacobs et al.  Taking into account these comparisons, we
estimate the uncertainty in the relevant rate coefficients of Mazzotta
et al.\ for DR onto \ion{Mg}{6}, \ion{Si}{8}, and \ion{S}{10} to be
+0\% and -55\%.

\subsubsection{Onto Oxygenlike \ion{Mg}{5}, \ion{Si}{7}, and \ion{S}{9}}
\label{subsubsec:O}

\citet{Jaco77b,Jaco79a} have published $LS$ rate coefficients for
\ion{Mg}{5}, \ion{Si}{7}, and \ion{S}{9}.  These rate coefficients were
fitted by \citet{Shul82a}. \citet{mazzotta98} use these results.  We
estimate the uncertainties in the DR rate coefficients for these ions
by comparing the rate coefficients for DR onto the isoelectronic
\ion{Fe}{19} for which a number of different calculations exist.
\citet{Jaco77a} and \citet{Rosz87c} published $LS$-coupling results.
\citet{Jaco77a} used the same technique as \citet{Jaco77b,Jaco79a}.
Their results were fitted by \citet{Shul82a} for use in ionization
balance calculations.  Hartree-Fock calculations with relativistic
corrections, using the code of \citet{Cowa81a}, were reported by
\citet{Dasg94a}.

In making our comparisons between the different calculations, we need
to take into account the relative importance of the different DR
channels for the ions of interest.  For \ion{Fe}{19}, near the
temperature of peak formation the $\Delta N=1$ $(N=2 \to N^\prime=3)$
channel appears to dominate the DR process \citep{Rosz87c,Dasg94a}.  It
is unclear how this extrapolates to the less highly charged
\ion{Mg}{5}, \ion{Si}{7}, and \ion{S}{9}.

For \ion{Fe}{19} at the temperature of peak formation, the rate
coefficients of \citet{Rosz87c} and \citet{Dasg94a} are, respectively,
factors of approximately 2.7 and 3.6 times larger than the result of
\citet{Jaco77a}.  Mazzotta et al.\ use the rates of
\citet{Jaco77b,Jaco79a} for \ion{Mg}{5}, \ion{Si}{7}, and \ion{S}{9}.
Based on our comparison for \ion{Fe}{19}, we estimate the uncertainty
in the high temperature rate coefficients for DR onto these ions to be
+260\% and -0\%.

\subsubsection{Onto Fluorinelike \ion{Si}{6} and \ion{S}{8}}
\label{subsubsec:F}

\citet{Jaco77b,Jaco79a} carried out $LS$-coupling calculations for
\ion{Si}{6} and \ion{S}{8}.  We are unaware of any other calculations
for these ions.  \citet{Shul82a} fit the data of Jacobs et al.
\citet{mazzotta98} use these fits.

We can estimate the important DR channels in \ion{Si}{6} and \ion{S}{8}
and the uncertainties in the relevant DR rate coefficients using the
theoretical results for the isoelectronic ions \ion{Fe}{18}.
\citet{Jaco77a}, using the same techniques as \citet{Jaco77b,Jaco79a},
have published an $LS$ rate coefficient.  \citet{Shul82a} presented a
fit to this rate coefficient for use in plasma modeling.
\citet{Rosz87b} calculated an $LS$ rate coefficient.  \citet{Dasg90a}
published Hartree-Fock calculations with relativistic corrections using
the code of \citet{Cowa81a}.  MCDF results were reported by
\citet{Chen88a}.  At the temperatures of peak formation for
\ion{Fe}{18}, the $\Delta N=1$ $(N=2 \to N^\prime=3)$ channel dominates
the DR process.  It is likely that the same situation exists for
\ion{Si}{6} and \ion{S}{8}.  At the temperature of peak formation for
\ion{Fe}{18}, the rate coefficients of \citet{Rosz87b},
\citet{Chen88a}, and \citet{Dasg90a} lie factors of approximately 3.5,
4.2, and 4.7, respectively, above the results of \citet{Jaco77a}.
Here, for \ion{Si}{6} and \ion{S}{8}, we use the recommended high
temperature DR rate coefficients of \citet{mazzotta98} and estimate the
uncertainty to be +370\% and -0\%.

\subsubsection{Onto Neonlike \ion{Si}{5}}
\label{subsubsec:Ne}

\citet{Jaco77b} and \citet{Roma88a} have calculated DR rate
coefficients using $LS$-coupling for \ion{Si}{5}.  At the peak in the
DR rate coefficient, the results of Jacobs et al.\ lie a factor of
$\approx 1.9$ below those of Romanik.  \citet{mazzotta98} use the
results of Romanik.  We use the theoretical rate coefficients
for DR onto \ion{Ar}{9} to estimate the uncertainty in the recommended
\ion{Si}{5} data.  $LS$ rate coefficients were given
by Romanik.  \citet{Chen86a} published MCDF rate coefficients.
HULLAC results were given by \citet{Four97a}.

The results of \citet{Roma88a} and \citet{Four97a} are in excellent
agreement.  The calculations of \citet{Chen86a} lie $\approx 15\%$
below those of Romanik and Fournier et al.  The close agreement between
these three different calculations and the large difference between the
results of \citet{Jaco77b} and Romanik for \ion{Si}{5}, strongly
suggests an error in the reported results of Jacobs et al.  This is
supported by a comparison of rate coefficients for DR onto \ion{Fe}{17}
by \citet{Arna92a}.  They found that the results of Chen and
Romanik were in good agreement but that the results of
\citet{Jaco77a}, calculated using the same techniques as
\citet{Jaco77b}, were a factor of $\approx 4.7$ times smaller.
Considering these various comparisons, we use the recommended rate
coefficient of \cite{mazzotta98} for \ion{Si}{5} and estimate the
uncertainty to be +0\% and -15\%.

\section{Solar and Stellar Upper Atmosphere Abundance Observations}

\subsection{Determining Relative Abundances}
\label{subsec:Determining}

Spectral line ratios can be used to determine relative abundances in
cosmic plasmas.  Photons are emitted in a spectral line at a rate per
second per steradian given by
\begin{equation}
R=
{A_{\rm ji}\over 8\pi}
\int_{\Delta V}
{n_{\rm j}\over n_{\rm q}}
{n_{\rm q}\over n_{\rm A}}
{n_{\rm A}\over n_{\rm H}}
{n_{\rm H}\over n_e}
n_{\rm e} dV
\label{eq:f}
\end{equation}
where
$R$ is the rate,
$A_{\rm ji}$ the radiative decay rate from excited ionic level $j$ to $i$,
$n_{\rm j}$ the number density of level $j$,
$n_{\rm q}$ the number density for the $q$-times charged ion in question,
$n_{\rm A}$ the number density for the element from which this ion is formed,
$n_{\rm H}$ the hydrogen number density,
$n_{\rm e}$ the electron number density, and
$\Delta V$ the emitting volume.
The radiation is emitted into $4\pi$ steradians.  For full disk spectra
it is usual to include a factor of 1/2 to account for photons emitted
towards the solar disk and absorbed by the photosphere.  At low
densities (i.e., in coronal plasma), we can write $n_{\rm j}/n_{\rm
q}=C_{\rm j}n_{\rm e}/A_{\rm ji}$ where $C_{\rm j}$ is the total
excitation rate coefficient of the excited level j by electron
collisions and we have assumed the branching ratio for the $j \to i$
radiative transition to be 1.  For the transitions of interest in the
solar FIP diagnostics under consideration here (i.e.  $n=2\rightarrow
n=2$ transitions in $L$-shell ions), excitation rate coefficients due
to other processes are negligible compared to those from electron
impact excitation.  Defining $f_q=n_{\rm q}/n_{\rm A}$ and changing the
integration variable from volume to temperature gives
\begin{equation}
\label{eq:R}
R={1\over 8\pi}{n_{\rm A}\over n_{\rm H}}
{n_{\rm H}\over n_e}\int _{\Delta T}C_{\rm j} f_{\rm q}
n_{\rm e}^2 {dV\over dT}dT
\end{equation}
where parameters not dependent on the electron temperature have been
taken outside the integral.  The quantity $n_{\rm e}^2 dV/dT$ is the
differential emission measure (DEM, also sometimes defined as $n_{\rm
e}^2 dV/d\log T$). The usual procedure in determining the DEM 
is to assume some functional form
(usually ${\rm DEM} \propto T^a$), evaluate $R$ for all lines, compare these
values with observations and iterate on the DEM distribution until
satisfactory agreement is obtained. This can be done for all the lines
of one particular element (usually Fe).  Then lines from other elements
can be analyzed the same way to determine relative element abundances.
However in this procedure the role of uncertainties in particular
atomic rate coefficients becomes obscured by the quantity of data and
number of iterations required.

A second technique for determining relative abundances is to use a line
from an ion of a specific element and another line from an ion of a
different element.  This is the technique which will most likely be
used to infer relative abundances from future solar satellite
missions.  The lines are chosen so that the respective values of
$C_{\rm j} f_{\rm q}$, to a first approximation, have similar
dependences on temperature, differing only by a multiplicative
constant.  Using Equation~\ref{eq:R} to take the ratio of the two
lines, it is easy to show that the ratio of the two integrals reduces
to this constant, which is determined from atomic parameters, times the
relative abundances.  In this approximation, the shape of the DEM has
no affect on the results.  This technique has the advantage of allowing
one to study how uncertainties in atomic physics affect the inferred
relative abundances.

\subsection{Solar Disk Observations}
\label{subsec:SolarDisk}

The first systematic analysis of coronal abundances over a wide
temperature region using high quality spectral data was probably that
of \citet{laming95} who used a variant of the first DEM method
described above.  They analyzed the full-disk solar corona spectrum of
\citet{malinovsky73} and covered a range of $\log
T=5.5-6.5$.  They found an abundance pattern consistent with that of
the slow speed solar wind, at least for solar coronal plasma with an
electron temperature $\log T > 5.8$ (i.e.  $T>9\times 10^5$~K). More
recent studies with lines from ions formed at similar high temperatures
corroborate this result \citep{laming99,white00}. But in what was a
surprise at the time, they found essentially no FIP enhancement in
transition region plasma at $\log T < 5.8$.  A similar result was found
earlier by \citet{noci88}, who used data from the Skylab S-055 EUV
spectroheliometer.

Interpreting spectra from these cooler temperatures poses a number
of problems.  This may be in part due to complications in the solar
physics at these temperatures \citep{feldman94}, or due to problems in
the interpretation of the data.  At transition region temperatures, the
solar emission measure $EM=n_e^2V\left(T\right)$ is an increasing
function of $T$.  The ``canonical'' quiet sun behaviour is $EM\propto
T^{1.5}$ \citep{jordan80}, though in certain solar features power laws
of $T^4$ or higher may be appropriate
\citep{cargill94,cargill97}. This is quite different from the situation
at $\log T > 6$ where in the quiet sun the emission
measure is in general much less steep, and may even be flat in certain
temperature regimes \cite[e.g.,][]{laming95}.
One consequence of the $EM$ temperature dependence is that in
observations of ions formed at $\log T  > 5.8$, where the emission
measure is not significantly temperature dependent, the detected line
emission comes primarily from regions near the temperatures of peak
fractional abundance for the observed ions. However, this is not the
case for ions with formation temperatures of $\log T < 5.8$.  For these
ions, the steeply rising $EM$ skews the temperature range over which
emission from the ions is sampled to temperatures well above those
where the ions peak in fractional abundance.  Under these conditions
uncertainties in the atomic data used to calculate the ionization
balance can significantly affect the inferred properties of the
observed plasma. In addition, at temperatures significantly above or
below those where an ion is formed in ionization equilibrium, we find
that the fractional ion abundances calculated by balancing the
electron-ion ionization and recombination rate coefficients are
extremely sensitive to errors in the atomic data.

Much of the work prior to that of \citet{laming95} had clearly observed
FIP effects at $\log T < 5.8$, principally using the
\ion{Mg}{6}/\ion{Ne}{6} and \ion{Ca}{9}/\ion{Ne}{7} intensity ratios in
the second, simpler method outlined in \S~\ref{subsec:Determining} for
determining relative abundances
\citep{feldman90,feldman93,widing89,widing92,widing93}.  One reason for
the apparent disagreement between these studies and that of
\citet{laming95} is probably that while Laming et al.\ studied a full
disk integrated spectrum, the prior work had concentrated on discrete
solar features observed by the Skylab SO-82A spectroheliograph.
Specifically, the Skylab investigations had concerned a coronal polar
plume \citep{widing92}, an impulsive flare \citep{feldman90}, an open
field active region \citep{widing93}, or a variety of such features
\citep{widing89}.

Interestingly the study of a coronal hole using a
\ion{Mg}{6}/\ion{Ne}{6} line ratio \citep{feldman93} yielded an
apparent FIP enhancement of about 2-2.5 at $\log T \approx 5.6$.  This
is slightly higher than the findings of \citet{laming95} for the full
disk sun in this lower temperature range.  However, given the estimated
uncertainties of 50\% in the \citet{feldman93} and \citet{laming95} in
this temperature range, and the different methods of analysis, there is
a suggestion that these results are in agreement.  The FIP enhancements
of \citet{feldman93} are also similar to or slightly higher than those
observed in the fast speed solar wind, which is believed to emanate
from coronal holes.  The results of \citet{feldman93} and
\citet{laming95} support the hypothesis that the FIP factor decreases
as one moves inward from the solar corona to the transition region.

A considerable amount of attention has been given to the open field
active region reported by \citet{widing93} and the coronal polar plume
reported by \citet{widing92}.  This is due to the observed FIP
enhancements being greater than an order of magnitude.  {\it SOHO}
observations of open field active regions using CDS have found results
\citep{young97} similar to those of the active region of
\citet{widing93}.  {\it SOHO} observations have also been carried out
of coronal polar plumes, principally using SUMER. \citet{doschek98}
made observations of the Si/Ne abundance ratio (as a proxy for the FIP
effect) and found a FIP enhancement of only a factor of 2. Other SUMER
observers find similar results \citep{delzanna99,young99}.  So far, it
has not been possible to reproduce the polar plume results of
\citet{widing92}.

\begin{figure}
\plotone{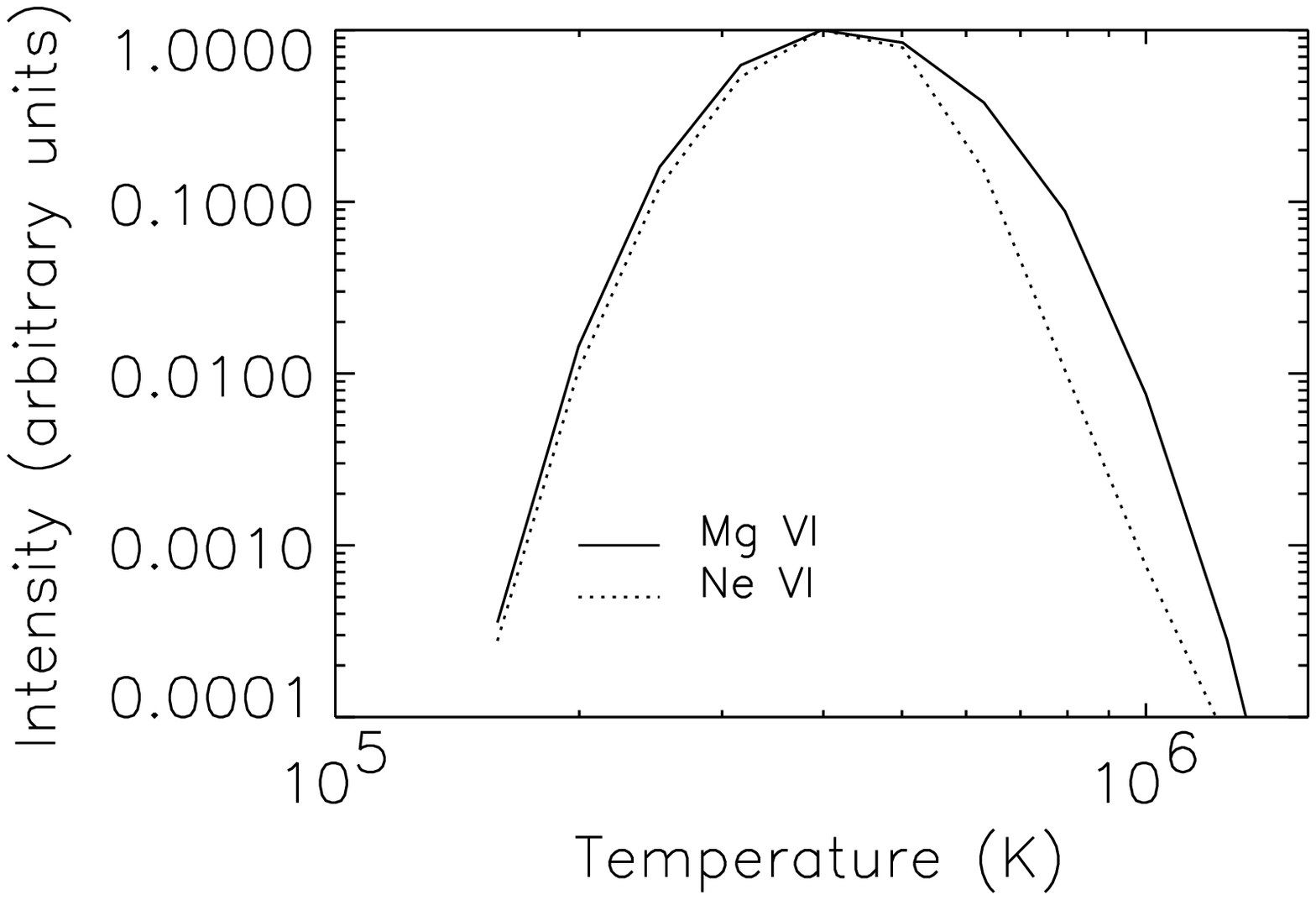}
\caption{Intensity curves for the emissivity (photons per second per
unit electron density per ion) times ionization fraction of the
\protect\ion{Mg}{6} 1190.09 and \protect\ion{Ne}{6} 558.59 \AA\  lines.
Both curves have been normalized to their maximum values.  The
correspondence between the two is almost exact for $T\le 5\times 10^5$
K. At higher temperatures, the \protect\ion{Mg}{6} line is stronger. A
steeply rising emission measure may weight emission from this
temperature region enough to produce large apparent Mg/Ne abundance
ratios, if this difference between the two curves is not accounted for
during data analysis.
\label{mgne}}
\end{figure}

The emission measure in the open field active region analyzed by
\citet{widing93} is relatively flat with temperature and the inferred
FIP enhancements of these features are believed to be reliable.
However, the emission measure plotted by \citet{widing92} for their
observed coronal polar plume rises very steeply with temperature.
Although the \ion{Mg}{6} and \ion{Ne}{6} ionization fractions match
each other very well near their temperature of peak formation, at
higher temperatures the \ion{Mg}{6} fraction becomes larger than that
for \ion{Ne}{6}. This is illustrated in Figure \ref{mgne} where the
relative intensities are plotted for the \ion{Mg}{6} 1190.09 and
\ion{Ne}{6} 558.59 \AA\ lines, using the ionization balance of
\citet{arnaud85}. The excitation rate coefficients are taken to vary as
$\exp\left(-\Delta E/kT\right)/\sqrt{T}$, where $\Delta E$ is the
excitation potential and $T$ is the electron temperature.  Under
conditions where the temperature structure of the plasma weights the
emission from these lines so that the temperature region greater than
$5\times 10^5$ is dominant, apparent abundance enhancements of Mg over
Ne may result simply due to the relative ionization fractions.
Realistic calculations by \citet{doschek00} for the emission measure
distributions found by \citet{widing92} suggest that an extra factor of
2-3 in the intensity ratio, on top of the usual FIP effect, could
result from this effect. Thus if the true FIP effect is a factor of
2-4, an apparent abundance enhancement of an order of magnitude or more
could result.

\subsection{Stellar Observations}

All observations of stellar coronae are necessarily disk
integrated.  In general in spectra of stellar coronae, individual line
ratios for FIP effect diagnostics are not as readily
available as they are in solar spectra. Most abundance results have
been derived using these spectra from emission measure plots, and
related techniques, where the effect of uncertainties in DR 
is much less transparent. However two papers that do use
specific line ratios are \citet{drake95} on Procyon and \citet{drake97}
on $\alpha$ Cen.   These sources are perhaps the two best observed
stellar coronae with the Extreme Ultraviolet Explorer ({\it EUVE}).

In general, analyses on {\it EUVE} spectra of solar-like stars have
found coronal abundance anomalies similar to the solar FIP effect. Only
Procyon appears to have a photospheric abundance corona. Coronal
abundances in active binary stars do not show such effects, but hint at
a trend of decreasing coronal metallicity with increasing activity.
These and other results prior to the launch of the {\it Chandra} and
{\it XMM-Newton} satellites are reviewed by \citet{feldman00}. With the
advent of high resolution stellar x-ray spectroscopy with {\it Chandra}
and {\it XMM-Newton}, the field of stellar coronae and their abundances
is set to become a rich area of research. Already new results have
raised the possibility of yet more varied coronal abundance patterns,
e.g., the inverse FIP effect \citep{brinkmann00} or enhanced Ne and Ar
\citep{drake01} in the coronae of HR 1099.

\section{Effects of DR Uncertainties on Relative Abundance Determinations}
\label{sec:Effects}

Our work focuses on studying particular line ratios because of their
importance for determining relative abundance ratios using data
collected from past, current, and future solar satellite missions.
These line ratios also offer a relatively transparent insight into the
effect of uncertainties in the DR rate coefficients on abundance
measurements.  {\it SOHO} studies have been carried out using
\ion{Mg}{6}/\ion{Ne}{6}, \ion{Mg}{7}/\ion{Ne}{7},
\ion{Mg}{9}/\ion{S}{9}, \ion{Mg}{9}/\ion{S}{10},
\ion{Si}{9}/\ion{S}{9}, and \ion{Si}{9}/\ion{S}{10} line ratios. 
Wavelengths for the observed lines are given
in detail in Table 3. An
additional abundance ratio that is of interest is \ion{Si}{10}/\ion{S}{10},
derived from the Si X 258.40 and 261.27 \AA\ lines and S X transitions at
259.50 and 264.24 \AA\,
which has been used in {\it EUVE} abundance studies on the coronae of
$\alpha$ Centauri \citep{drake97}.  We consider all of the 7 above
listed line ratios in our discussion below.

\subsection{Ionization Balance Calculations}
\label{subsec:Ionization}

The ionization fraction $f_q$ of the ion with charge $q$ is given by
\begin{equation}
{df_q\over dt} =
n_e\left[C_{ion,q-1}f_{q-1}-C_{ion,q}f_q +
\left(C_{rr,q+1} + C_{dr,q+1}\right)f_{q+1} -
\left(C_{rr,q}+ C_{dr,q}\right)f_q\right]
\label{eqn1}
\end{equation}
where $C_{ion,q}$, $C_{rr,q}$, and $C_{dr,q}$ are the rate coefficients
for electron impact ionization, RR, and DR respectively, out of the
charge state $q$. For the coronal plasmas under consideration here
three-body recombination
can be safely neglected. In ionization equilibrium, $df_q/dt=0$ which
gives $Z$ independent linear equations for the $Z+1$ charge states. The
constraint that all charge state fractions must add up to unity
supplies the final equation, allowing us to solve the $Z+1$ linear
equations by LU decomposition \citep[see][for details]{press92} of the
matrix formed by the right hand side of Equation~\ref{eqn1}.  This set
of linear equations is solved repeatedly with rate coefficients
appropriate to different electron temperatures to find the temperature
dependence of the different charge state fractions.  Initially the
ionization, RR adn DR rate coefficients
are taken from the same sources as used by \citet{mazzotta98}.

\subsection{Relative Abundances}
\label{subsec:Relative}

Using Equation~\ref{eq:R}, the relative abundances for
two elements, A1 and A2 is given by
\begin{equation}
\label{eq:nA1vnA2}
{n_{\rm A1} \over n_{\rm A2}}=
{R_{\rm A1} \over  R_{\rm A2}} \times
{\int _{\Delta T} C_{\rm j2} f^{\rm A2}_{\rm q} n_{\rm e}^2 {dV\over dT} dT
\over
 \int _{\Delta T} C_{\rm j1} f^{\rm A1}_{\rm q} n_{\rm e}^2 {dV\over dT} dT}.
\end{equation}
This equation is first evaluated with ionization balance calculations
which use the unscaled DR rate coefficients of \citet{mazzotta98}.  To
investigate the effects that the various sets of DR rate coefficients
have on the inferred value of $n_{\rm A1}/n_{\rm A2}$, we define the
quantity
\begin{equation}
\label{eq:S}
S=
\Biggl({
\int _{\Delta T} C_{\rm j2} f^{\rm A2}_{\rm q} n_{\rm e}^2 {dV\over dT} dT
\over
\int _{\Delta T} C_{\rm j1} f^{\rm A1}_{\rm q} n_{\rm e}^2 {dV\over dT} dT}
\Biggr)_{\rm new}
\Biggl/
\Biggl({
\int _{\Delta T} C_{\rm j2} f^{\rm A2}_{\rm q} n_{\rm e}^2 {dV\over dT} dT
\over
\int _{\Delta T} C_{\rm j1} f^{\rm A1}_{\rm q} n_{\rm e}^2 {dV\over dT} dT}
\Biggr)_{\rm old}
\end{equation}
where the subscripts old and new refer, respectively, to ionization
balance calculations carried out using the unscaled and scaled DR rate
coefficients of \citet{mazzotta98}.  Multiplying the right-hand-side of
Equation~\ref{eq:nA1vnA2} by $S$ yields the new inferred relative
abundances.  We note that a decrease(increase) in the calculated value
of $S$ corresponds to a decrease(increase) in the inferred FIP
enhancement for a given value of $R_{\rm A1}/R_{\rm A2}$.

\subsection{Effects}
\label{subsec:Effects}

For our calculations of $S$ we made a number of simplifying
assumptions.  For the pairs of lines we are concerned with here, to a
first approximation the values of $C_{\rm j}f_{\rm q}$ have the same
temperature dependence.  To simplify the calculations we have assumed a
flat DEM ($\propto T^0$).  We further simplify the calculations by
neglecting the temperature dependence of $C_{\rm j}$, setting it equal
to 1 for each ion.  In this way we are able to focus specifically on
how the uncertainty in the DR rate coefficients affects $S$.  We then
calculated $S$ for all 1728 sets of DR rate coefficients.  The
calculations were carried out over the temperature range $\Delta T$
from $\log T=5.0 -7.0$.  This covers the temperature range over which
the ions of interest form.  For brevity, we have listed in
Table~\ref{tab:Srange} only the range of $S$ values for all 1728
variations.  Clearly the uncertainties in the DR rate coefficients can
have a dramatic affect on any inferred FIP enhancement factors.  We
note that Table~\ref{tab:Srange} does not display the correlations
between the values of $S$ for different line ratios for each set of DR
variations.

We have reduced the set of DR variations using our current hypothesis
for the structure of the FIP effect.  \ion{Mg}{7} and \ion{Ne}{7} are
transition region temperature ions which have been observed by
\citet{laming99} from close to the solar limb out into the corona
(where these ions are observed far away in temperature from where they
peak in abundance).  The inferred FIP effect is expected to be $\approx
4$ in the coronal observations and decrease as one moves to transition
region temperatures closer to the solar limb.  This expected behavior
for the Mg/Ne abundance ration can be seen in Table~\ref{tab:fip} which
is reproduced here from \citet{laming99}.  Line emission from
\ion{Mg}{9} and \ion{S}{10} is always dominated by coronal temperature
plasmas and the \ion{Mg}{9}/\ion{S}{10} ratio is expected to display a
FIP factor of 4 at all positions listed in Table~\ref{tab:fip}.
\citet{laming99} measured a FIP factor of $\approx 3$.  For our
selection criteria, we take only those DR variations for which the FIP
effect inferred using \ion{Mg}{7}/\ion{Ne}{7} line ratio varies by less
than $\pm 25\%$ and which also increase the FIP effect inferred using
\ion{Mg}{9}/\ion{S}{10}.  The 25\% is based on the estimated errors in
the measured line ratios.  We found 548 sets which met these
constraints.  This set was reduced to 274 by taking into account the
fact that changes in the DR rate coefficient onto neonlike \ion{Si}{5}
had essentially no affect on the results.  We have
listed in Table~\ref{tab:Srange} the range of $S$ for this reduced set
of variations.  Clearly, even in this reduced set the uncertainties in
the DR rate coefficients can still have a dramatic affect on many
inferred FIP enhancement factors.

\subsection{Solar Off-Limb Observations: A Further Test}

At heights of 50 arcsec or more from the solar limb, the corona becomes
essentially isothermal. As one away from the solar limb, the amount of
cooler transition region plasma decreases quickly with height, while
the hotter coronal gas diminishes much less rapidly, and quickly
becomes the dominant component. A striking illustration of this in the
case of a solar equatorial streamer is given in \citet{feldman99}. In
Figure 3 of their paper they plot the variation of intensity with
distance from the solar limb for various transitions from ions of
\ion{Si}{7}-\ion{Si}{12}.  All line intensities $I$ show the same slope
$d\log I/dr$ and hence the same temperature, which from the slope
evaluates to about $1.5\times 10^6$ K.  Feldman et al.\ plot in their
Figure 4 the loci of emission measures determined from these Si ions.
Possible calibration uncertainties exist for the \ion{Si}{12} results
as it is at the extreme short wavelength end of the SUMER second order
bandpass \citep{laming97}.  Ignoring the \ion{Si}{12} loci, the
temperature determined from the intersection of the remaining emission
measure loci is $\log T = 6.11\pm 0.04$. This is similar to that
determined from the height dependence of the emission in the various
lines, assumed to fall off according to hydrostatic equilibrium.

\begin{figure}
\plotone{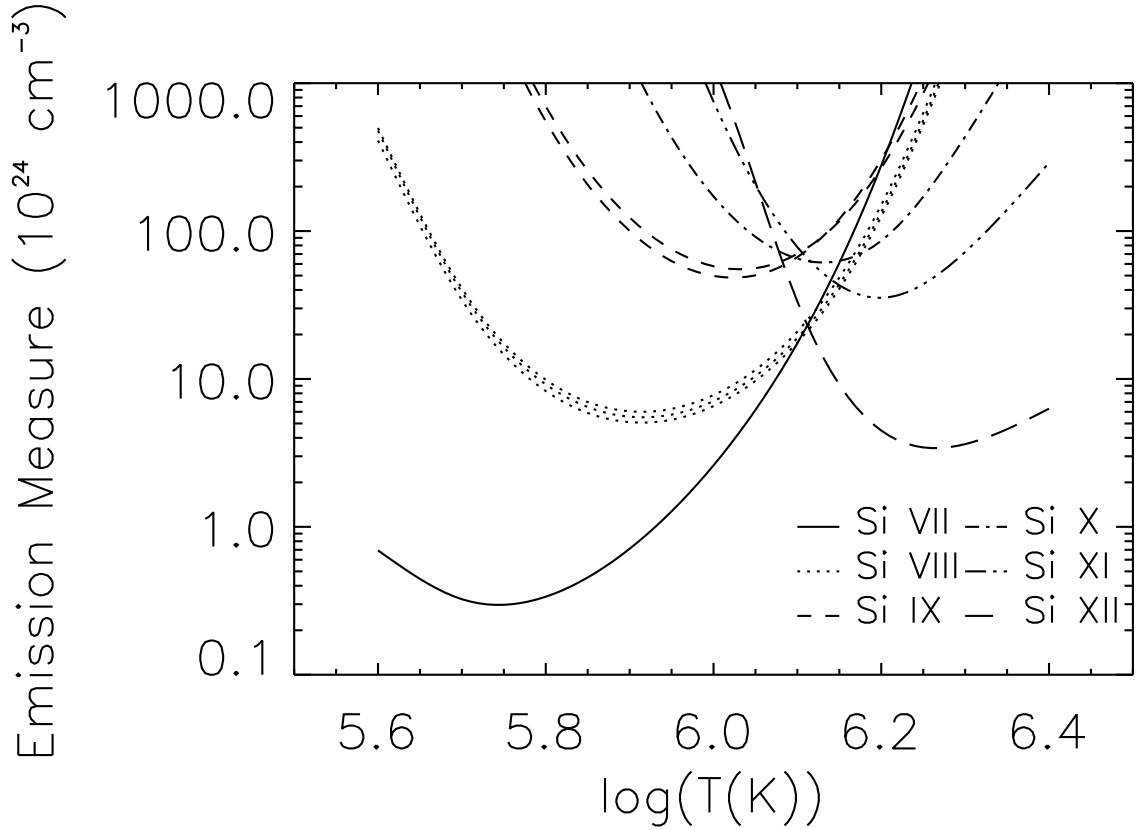}
\caption{Emission measure loci for the ions \protect\ion{Si}{7}-
\protect\ion{Si}{12} determined from SUMER observations of an
equatorial streamer using the ionization balance calculations of
\protect\citet{mazzotta98}. The curves intersect at a temperature close
to $\log T = 6.15$, suggesting the conclusion that the plasma is
isothermal. The spectral lines considered for each ion are:
\protect\ion{Si}{7} 1049.22 {\AA}; \protect\ion{Si}{8} 944.38, 949.22,
and 1445.75 {\AA}; \protect\ion{Si}{9} 950.14 and 694.70 {\AA};
\protect\ion{Si}{10} 638.94 {\AA}; \protect\ion{Si}{11} 580.91 {\AA};
and \protect\ion{Si}{12} 499.41 {\AA}.  These give three curves for
\protect\ion{Si}{8}, two for \protect\ion{Si}{9} and one for each
remaining ion.
\label{fig_si0}}
\end{figure}

\citet{feldman99} used the older ionization balance of
\citet{arnaud85}.    In order to assess the impact of our suggested
changes to dielectronic recombination rate coefficients, we have
remeasured the line intensities and replotted the emission measure
determined using the ionization balance of \citet{mazzotta98}.  These
results are shown in Figure \ref{fig_si0}.  Comparing with Figure 4 of
\citet{feldman99} we can see that the temperature at which the various
curves intersect is now given by $\log T=6.14\pm 0.05$.

\begin{figure}
\plotone{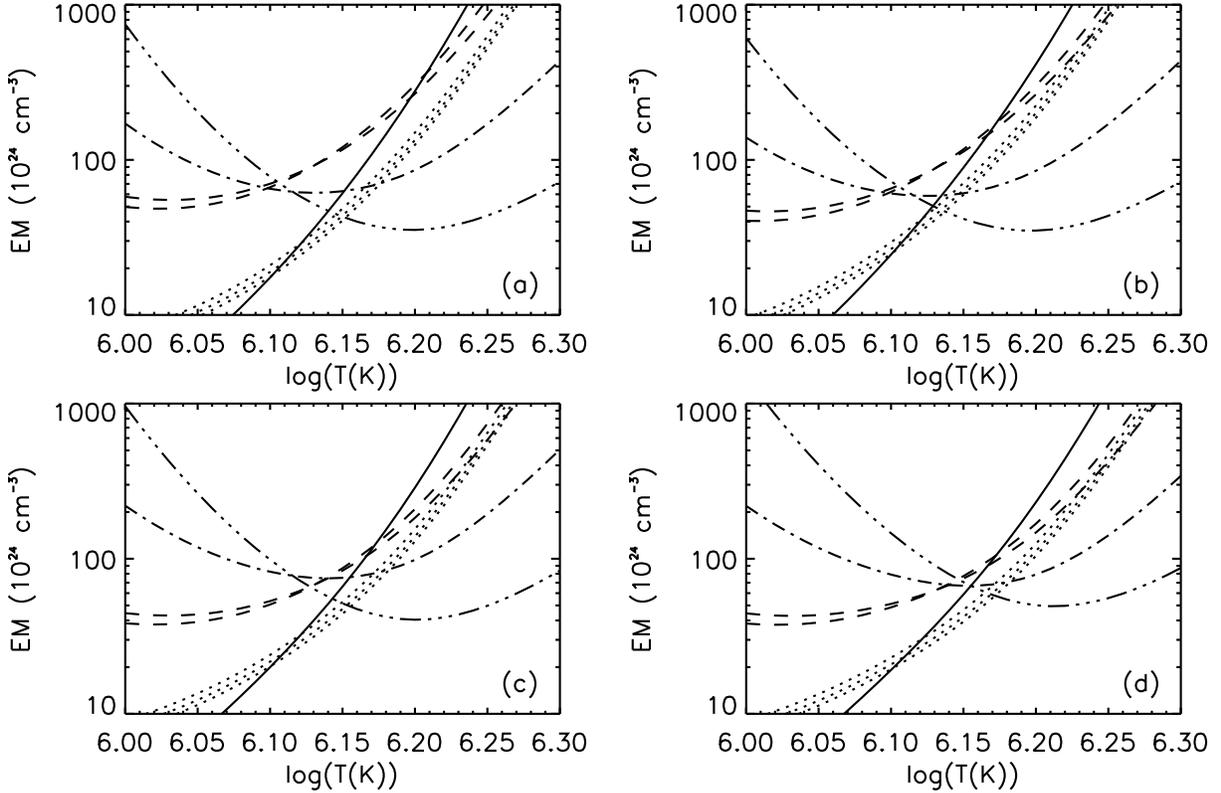}
\caption{Emission measure (EM) loci from Figure~\ref{fig_si0} for the
temperature region $6.0 \le \log T \le 6.3$.  Figure~3(a) shows the
results using the \protect\citet{mazzotta98} ionization balance.
Figures~3(b), 3(c), and 3(d) show, respectively, the plots
corresponding to variations 4, 2, and 3 of the Mazzotta et
al.\ ionization balance.  Variations 4, 5, 7, and 9 are very similar as
are variations 1, 2, 6 and 8.  As can be seen from the Figure,
variation 3 appears to give the best improvement to the inferred (and
expected) isothermal nature of the observed coronal plasma, followed by
variation 2, with variation 4 being the least successful. The curves
due to \protect\ion{Si}{12} 499.41 \AA\ have been omitted from these
plots due to suspicions about the instrument calibration.
\label{fig_si1}}
\end{figure}

We have investigated the effect of varying the DR rate coefficients in
the Si ionization balance calculations.  We have looked at variations
of the rate coefficients in a ``trial and error'' fashion.  Nine
plausible sets of variations are given in Table
\ref{tbl1}.  Others exist among our 1728 attempts, but these nine serve
to illustrate one effect we are interested in. In Figure
\ref{fig_si1}(a) we replot the temperature region $6.0 < \log T < 6.3$
from the previous plot.  Figures~\ref{fig_si1}(b), (c), and (d) give,
respectively, the plots corresponding to variations 4, 2, and 3 of the
\citet{mazzotta98} DR rate coefficients.  Variations 4, 5, 7, and 9 are
very similar as are variations 1, 2, 6 and 8.  The curves due to
\ion{Si}{12} 499.41 \AA\ have been omitted from these plots due to
suspicions about the instrument calibration.  In all panels the degree
of overlap between the various intersections of the emission measure
loci has improved.  The improvement is best for variation 3, followed
by variation 2.  Variation 4 is the least successful.  Thus the
important conclusion that we are observing an isothermal plasma is, if
anything, strengthened by our considerations of these plausible
variations in the DR rate coefficients.

\section{Conclusions}

We have described on the basis of atomic physics theory what uncertainties
may be present in the DR rate coefficients commonly used in ionization
balance calculations.  Of course this situation will only be definitely
improved by further calculations and experiments, but given the number
of individual rate coefficients required for astrophysical modeling
purposes, this is an enormous amount of work.  To help prioritize the
needed atomic data, and to provide further motivation, we have
investigated how the uncertainties in the DR rate coefficients
translate into uncertainties in inferred FIP factors.  We find that,
depending on the specific line ratio, inferred FIP factors can be a
factor of 5 smaller or 1.6 times larger that the FIP factor inferred
using the unscaled DR rate coefficients of \citet{mazzotta98}.

Taking the unconstrained data in Table~\ref{tab:Srange} and using the
ratio of the maximum over minimum value of $S$ for each line ratio, we
can prioritize the need for DR measurements and calculations of the
various isoelectronic sequences studied here.  Listing the various line
ratios in decreasing order of $S_{max}/S_{min}$ yields:
\ion{Si}{9}/\ion{S}{9}, \ion{Mg}{9}/\ion{S}{9},
\ion{Si}{10}/\ion{S}{10}, \ion{Mg}{9}/\ion{S}{10},
\ion{Si}{9}/\ion{S}{10}, \ion{Mg}{6}/\ion{Ne}{6}, and
\ion{Mg}{7}/\ion{Ne}{7}.  Based on the order and frequency with which
the different isoelectronic sequences occur in this list, we prioritize
our list of needed rate coefficients for DR onto specific isoelectronic
sequences, in decreasing order of importance, as follows: O-like,
C-like, Be-like, N-like, B-like.  To this we append F-like, Li-like,
He-like, and Ne-like based on the range of the DR scale factors given
in Table~\ref{tbl1}.

We have arrived at our list through admittedly subjective means, but we
believe that in this case the ends justify the means.  Faced with the
current degree of uncertainty in the dielectronic recombination rate
coefficients relevant to astrophysical plasmas, the number of required
calculations and measurements is a daunting task.  Our aim in this work
has been to prioritize this work, and to point out those ions where the
uncertainties in the DR rate coefficients has the most impact on the
analysis of astrophysical UV and X-ray spectra.

\acknowledgements

We thank to Arati Dasgupta, Verne Jacobs, Pasquale Mazzotta, Kengo
Moribayashi, and Takako Kato for stimulating conversations and
correspondence.  DWS was supported in part by NASA Solar Physics
Research, Analysis, and Suborbital Program grant NAG5-9581 to Columbia
University.  JML was supported by the NRL/ONR Solar Magnetism and the
Earth's Environment 6.1 Research Option and by NASA Contracts W19473,
W19539 and the 2000 SEC Guest Investigator Program.
The SUMER project is financially supported by DARA, CNES,
NASA, and the ESA PRODEX program (Swiss Contribution). SUMER is a part
of SOHO, the Solar and Heliospheric Observatory, of ESA and NASA.

\vfill
\eject

\vfill
\eject

\clearpage

\begin{deluxetable}{clccccc}
\tablecaption{Scale factors for the recommended DR rate coefficients of
\protect\citet{mazzotta98} onto selected ions of Ne, Mg, Si, and S.
\label{tab:scalefactors}}
\tablewidth{0pt}
\tablehead{
\colhead{Isoelectronic} & \colhead{Ions} && \multicolumn{4}{c}{DR Scale
Factors}\\
\colhead{Sequence} & \colhead{ } && \multicolumn{4}{c}{}}
\startdata
He&\ion{Ne}{9},              \ion{Si}{13}            &&  -  & 0.8 & 1.0 &
1.2 \\
Li&\ion{Ne}{8}, \ion{Mg}{10}, \ion{Si}{12}           &&  -  & 0.8 & 1.0 &
1.2 \\
Be&\ion{Ne}{7}, \ion{Mg}{9}, \ion{Si}{11}            &&  -  &  -  & 1.0 &
1.6 \\
B &\ion{Ne}{6}, \ion{Mg}{8}, \ion{Si}{10}            &&  -  &  -  & 1.0 &
1.7 \\
C &\ion{Ne}{5}, \ion{Mg}{7}, \ion{Si}{9}, \ion{S}{11}&& .31 & .62 & 1.0 &  -
\\
N &             \ion{Mg}{6}, \ion{Si}{8}, \ion{S}{10}&&  -  & .45 & 1.0 &  -
\\
O &             \ion{Mg}{5}, \ion{Si}{7}, \ion{S}{9} &&  -  &  -  & 1.0 &
3.6 \\
F &                          \ion{Si}{6}, \ion{S}{8} &&  -  &  -  & 1.0 &
4.7 \\
Ne&                          \ion{Si}{5}             &&  -  & .85 & 1.0 &  -
\\
\enddata
\end{deluxetable}

\begin{deluxetable}{lcccccc}
\tablecaption{Range of scale factor $S$ for all 1728 sets of DR
variations (unconstrained) and for the reduced set of 274 variations
(constrained).
\label{tab:Srange}}
\tablewidth{0pt}
\tablehead{
\colhead{Line ratio} && \multicolumn{5}{c}{Scale Factor $S$} \\
\cline{3-7}
\colhead{ }          && \multicolumn{2}{c}{Unconstrained}
&                     & \multicolumn{2}{c}{Constrained} \\
\cline{3-4} \cline{6-7}
\colhead{ }          && \colhead{Minimum} & \colhead{Maximum}
&                     & \colhead{Minimum} & \colhead{Maximum}}
\startdata
\ion{Mg}{6}/\ion{Ne}{6}  && 0.60 & 1.11 && 0.62 & 1.11 \\
\ion{Mg}{7}/\ion{Ne}{7}  && 0.67 & 1.22 && 0.75 & 1.22 \\
\ion{Mg}{9}/\ion{S}{9}   && 0.33 & 1.29 && 0.41 & 1.29 \\
\ion{Mg}{9}/\ion{S}{10}  && 0.51 & 1.64 && 1.00 & 1.64 \\
\ion{Si}{9}/\ion{S}{9}   && 0.20 & 1.01 && 0.24 & 1.00 \\
\ion{Si}{9}/\ion{S}{10}  && 0.36 & 1.14 && 0.43 & 1.14 \\
\ion{Si}{10}/\ion{S}{10} && 0.43 & 1.60 && 0.71 & 1.60 \\
\enddata
\end{deluxetable}

\begin{deluxetable}{lcccc}
\tablecaption{Observed Coronal FIP Fractionations 
\protect\citep[from][]{laming99}
\label{tab:fip}}
\tablewidth{0pt}
\tablehead{
\colhead{Line ratio} & \colhead{position 1\tablenotemark{a}}
& \colhead{position 2}& \colhead{position 3}
& \colhead{position 4}}
\startdata
\ion{Mg}{6} 1190.09/\ion{Ne}{6} 558.59 & $\ldots$    & $7.4\pm1.4$ &
$4.5\pm0.8$ & $3.5\pm0.8$\\
\ion{Mg}{7} 868.11/\ion{Ne}{7} 895.17  & $2.0\pm0.4$ & $2.6\pm0.5$ &
$4.1\pm0.7$ & $4.1\pm0.7$\\
\ion{Mg}{9} 749.55/\ion{S}{9} 871.71    & $3.6\pm0.8$ & $3.5\pm0.4$ &
$3.2\pm0.1$ & $3.1\pm0.1$\\
\ion{Mg}{9} 749.55/\ion{S}{10} 776.37   & $3.0\pm1.1$ & $3.2\pm1.0$ &
$2.6\pm0.8$ & $3.0\pm0.9$\\
\ion{Si}{9} 950.14/\ion{S}{9} 871.71   & $4.8\pm2.4$ & $3.6\pm0.5$ &
$2.7\pm0.1$ & $2.8\pm0.1$\\
\ion{Si}{9} 950.14/\ion{S}{10} 776.37  & $4.6\pm1.7$ & $3.0\pm0.9$ &
$2.0\pm0.7$ & $2.4\pm0.8$\\
\enddata
\tablenotetext{a}{Position 1 corresponds to a slit position covering
14'' in radial distance over the solar limb. Positions 2, 3, and 4 are
successively 14'' further out in radial distance from the solar limb.}
\end{deluxetable}

\begin{deluxetable}{cccccccccc}
\tablecaption{Scale factors for initial ions by which the recommended
DR rate coefficients of \protect\citet{mazzotta98}
have been multiplied by for
the 9 variations selected in a ``trial and error'' fashion.
\label{tbl1}}
\tablewidth{0pt}
\tablehead{
\colhead{Variation} & \multicolumn{9}{c}{Isoelectronic Sequence} \\
\colhead{Number}
& \colhead{He}
& \colhead{Li}
& \colhead{Be}
& \colhead{B}
& \colhead{C}
& \colhead{N}
& \colhead{O}
& \colhead{F}
& \colhead{Ne}}
\startdata
1 & 1.0 & 0.8 & 1.0 & 1.7 & 0.62 & 1.0 & 1.0 & 1.0 & 1.0\\
2 & 1.0 & 0.8 & 1.0 & 1.7 & 0.62 & 1.0 & 1.0 & 4.7 & 1.0\\
3 & 1.0 & 0.8 & 1.6 & 1.7 & 0.62 & 1.0 & 1.0 & 4.7 & 1.0\\
4 & 1.0 & 1.0 & 1.0 & 1.0 & 0.62 & 1.0 & 1.0 & 1.0 & 1.0\\
5 & 1.0 & 1.0 & 1.0 & 1.0 & 0.62 & 1.0 & 1.0 & 4.7 & 1.0\\
6 & 1.2 & 0.8 & 1.0 & 1.7 & 0.62 & 1.0 & 1.0 & 1.0 & 1.0\\
7 & 1.2 & 1.0 & 1.0 & 1.0 & 0.62 & 1.0 & 1.0 & 1.0 & 1.0\\
8 & 1.2 & 1.0 & 1.0 & 1.7 & 0.62 & 1.0 & 1.0 & 1.0 & 1.0\\
9 & 1.2 & 1.2 & 1.0 & 1.0 & 0.62 & 1.0 & 1.0 & 1.0 & 1.0\\
\enddata
\end{deluxetable}

\end{document}